\documentclass[twocolumn,prb,showpacs,superscriptaddress]{revtex4}

\usepackage{graphicx}
\usepackage{rotating}
\usepackage{amsmath}
\usepackage{amsfonts}
\usepackage{amssymb}
\usepackage{enumerate}
\usepackage{longtable}
\usepackage{dcolumn}
\usepackage{bm}
\usepackage{color}

\begin{document}
\newcommand{\cred}{\color{red}}
\title{Single-particle versus pair condensation of hard-core bosons with correlated hopping}

\author{K.P. Schmidt}
\email{kaiphillip.schmidt@epfl.ch}
\affiliation{Institute of Theoretical Physics, \'{E}cole Polytechnique F\'{e}d\'{e}rale de Lausanne, CH 1015 Lausanne, Switzerland}
\author{J. Dorier}
\affiliation{Institute of Theoretical Physics, \'{E}cole Polytechnique F\'{e}d\'{e}rale de Lausanne, CH 1015 Lausanne, Switzerland}
\author{A. L\"auchli}
\affiliation{Institut Romand de Recherche Num\'erique en Physique des Mat\'eriaux (IRRMA), CH-1015 Lausanne, Switzerland}
\author{F. Mila}
\affiliation{Institute of Theoretical Physics, \'{E}cole Polytechnique F\'{e}d\'{e}rale de Lausanne, CH 1015 Lausanne, Switzerland}
\date{\rm\today}

\begin{abstract} 
We investigate the consequences of correlated hopping on 
the ground state properties of hard-core bosons on a square lattice as revealed by extensive
exact diagonalizations and quantum Monte Carlo simulations.  While for non interacting 
hard-core bosons the effective attraction induced by the correlated hopping leads 
to phase separation at low density, we show that a modest  nearest-neighbor
repulsion suppresses phase separation, leading to a remarkable
low-density pairing phase with no single particle Bose-Einstein condensation 
but long-range two-particle correlations, signaling a condensation of pairs.
We also explain why the unusual properties of the pairing phase are a real challenge for standard one-worm quantum Monte Carlo simulations.
\end{abstract}


\maketitle

\section{Introduction}
Models of interacting bosons arise naturally in different fields of condensed matter and of atomic physics. The interplay of kinetics and interaction leads to interesting physical phases like superfluids (SF), Mott insulators (MI) or supersolids. Phase separation (PS) is also a common feature in bosonic systems. Usually, the nearest-neighbor hopping and local repulsions are the dominant processes. In frustrated physical systems however, 
correlated hopping can be the dominant source of kinetic energy. In a mean-field analysis, it has been shown that such purely kinetic terms favor the existence of a new pairing phase which shows a molecular condensate ($\langle b^\dagger_i b^\dagger_j\rangle\neq 0$) without single-particle condensation $(\langle b^\dagger_i \rangle=0)$\cite{bendj05}. Similar phases have been discussed in atomic physics in the context of ultracold atoms\cite{kagan02,kuklo04,roman04,radzi04,polle06}. In particular, a continuous transition from an atomic to a molecular condensate is of interest since it would be a realization of an Ising transition between two different superfluid phases. This quantum phase transition has to be contrasted with the smooth BEC-BCS crossover as e.g. in an atomic Fermi gas. Here both phases can be distinguished by their symmetry. In a standard superfluid the $U(1)$ symmetry is completely broken while in a paired superfluid a residual discrete $Z_2$ symmetry remains. Evidence beyond mean-field for the existence of such a paired superfluid (PSF) is still lacking however, and its detailed physical properties and the quantum phase transitions out of such a phase remain to be understood.

In this work we study numerically by exact diagonalization (ED) and quantum Monte Carlo (QMC) the physical properties of this exotic pairing phase of bosons. We first give an introduction to the model. In Sect.~\ref{ED} we discuss the ED results, including the global phase diagram of the model, the low-density limit, and the relevant correlation functions. Next we focus on the determination of the physical properties for larger system sizes obtained by QMC (Sect.~\ref{QMC}). At the end we collect the major findings and give some conclusions.
\section{Model}
We consider a model of hard-core bosons with correlated hopping and nearest-neighbor repulsion on the two-dimensional square lattice
\begin{eqnarray}
 H&=&-t\sum_i\sum_{\delta=\pm x,\pm y} b^\dagger_{i+\delta}b^{\phantom{\dagger}}_i-\mu\sum_i n_i\\\nonumber
&&-t^\prime\sum_i\sum_{\delta=\pm x;\delta^\prime=\pm y} n_i\left[b^\dagger_{i+\delta}b^{\phantom{\dagger}}_{i+\delta^\prime}+{\it h.c.}\right]\\\nonumber
&&+V\sum_{i}\sum_{\delta=+x,+y}n_{i+\delta}n_i
\end{eqnarray}
where $n_i=b^\dagger_i b^{\phantom{\dagger}}_i$ is the boson density at site $i$, $\mu$ is the chemical potential, $t$ is the nearest-neighbor hopping amplitude, $t^\prime$ denotes the strength of the correlated hopping, and $V$ is a nearest-neighbor repulsion. In the following we set $t+t'=1$ and plot the results as a function
of $t'$ only. We treat the bosons as hard-core particles, i.e. the possible local states are restricted to $n_i\in\{0,1\}$.

For vanishing $t^\prime$, the model can be mapped onto a XXZ model in a magnetic field. The phase diagram of the model is rich and rather well understood\cite{scale95,batro00,heber02,schmi02}. At half filling, one has a SF for $V<2t$ and a checkerboard solid for $V>2t$. For $V=2t$ the model maps to the Heisenberg model. Away from half filling, a large negative or positive chemical potential leads either to an empty or a full system. The transition between the checkerboard solid and the SF is likely to be first order except at the Heisenberg point at half filling. A supersolid phase is not realized for this model~\cite{batro00}.

For finite correlated hopping, the situation is much less clear. In this work we are not interested in possible charge ordering phases resulting from the nearest-neighbor interaction. We focus on the role of the correlated hopping for the realization of unconventional SF phases. For vanishing repulsion, mean field theory targeting a homogeneous solution predicts a phase transition between a standard SF with single-boson condensation and an exotic PSF with boson-pair condensation for large enough correlated hopping\cite{bendj05}. Indeed, the basic physical effect of the correlated hopping is to act as a binding mechanism for bosons. This however opens the possibility that the system undergoes PS, similar to bosonic ring-exchange models~\cite{rouss05}.
We will show in this work that this is indeed the case. The role of the nearest-neighbor repulsion is therefore crucial. We propose the following scenario. The system displays PS if the nearest-neighbor repulsion is small. On intermediate scales of the repulsion the PS is significantly reduced and the boson-pair condensation region is stabilized. Note that this is different for bosonic ring-exchange models where a finite repulsion only gives rise to a standard SF\cite{rouss05b}. For large repulsion the pairs themselves get broken up and a standard SF is recovered.

\section{Exact diagonalization}
\label{ED}

In this section we discuss results obtained by ED on finite clusters. The phase diagram has been obtained on systems with up to $N_{\rm s}=36$ sites. The low-density limit (up to 4 particles) was calculated up to a cluster with $N_{\rm s}=58$. In the following we will first discuss the full phase diagram, then the low-density limit and finally relevant correlation functions.
\begin{figure}
    \begin{center}
        \includegraphics[width=\columnwidth]{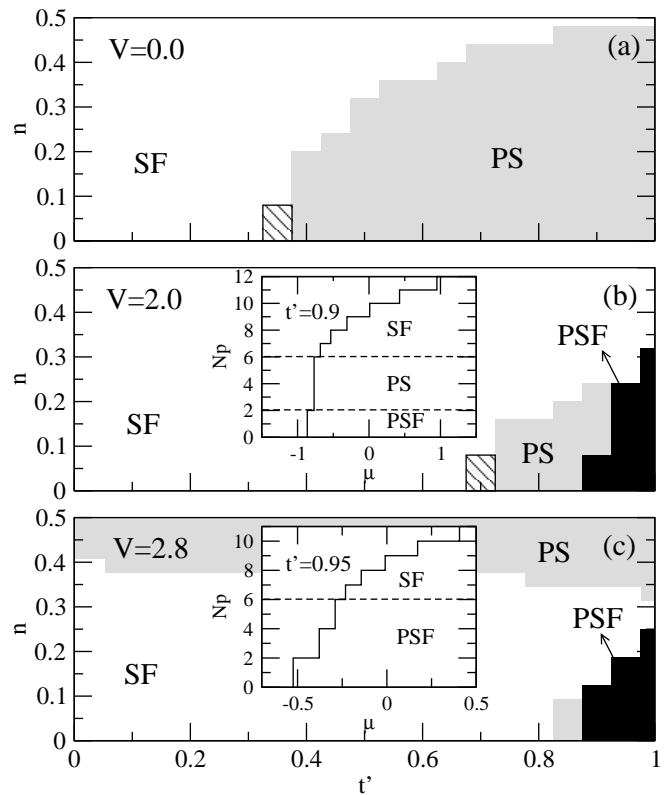}
    \end{center}
    \caption{Phase diagram obtained by ED. (a) $V=0$ and $N_s=25$ (b) $V=2.0$ and $N_s=25$ (c) $V=2.8$ and $N_s=32$. White represents the SF phase, grey denotes PS, and black represents the PSF. The tiny shaded region marks a region of pairing which is only present for a finite size system. {\it Insets}: Total particle number $N_{\rm p}$ as a function of $\mu$ for $t^\prime=0.95$.}
    \label{pd}
\end{figure}

The phase diagram of the model is shown in Fig.~\ref{pd} as a function of $t^\prime$ and density $n$ for different values of the nearest-neighbor repulsion. White regions represent SF regions, grey regions PS, and black regions the PSF. The different phases have been determined in the following way. Depending on the chemical potential, the system is filled by a certain number of particles. In a conventional phase the number of particles increases by one when the chemical potential is varied. In the PSF the insertion of a single boson is suppressed due to the presence of a pairing gap, and pairs of bosons enter the system. In the case of PS, boson-pairs are unstable towards cluster formation and the density jumps by a finite amount when the chemical potential is changed. The three different phases SF, PSF, and PS can then be characterized by the way in which the total particle number $N_{\rm p}=nN_{\rm s}$ changes upon changing the chemical potential with $\Delta (N_{\rm p})=1$, $\Delta (N_{\rm p})=2$, or $\Delta (N_{\rm p})>2$ respectively. Typical curves of the total number of particles $N_{\rm p}$ as a function of $\mu$ are shown in the insets of Fig.~\ref{pd}.

Let us look first at $V=0$. The phase diagram is shown in Fig.~\ref{pd}a for the $N_{\rm s}=25$ cluster. For not too large values of $t^\prime$ or for large densities a SF phase is realized. The SF becomes unstable to PS at lower densities for large correlated hopping. The tiny shaded region at the left edge of the PS at low densities (also present in Fig.~\ref{pd}b) represents a system with exactly two bosons which are paired by the correlated hopping. It does not present a true low density phase. Practically, the PSF is absent for this case. One therefore finds that the correlated hopping binds the bosons so strongly that the system phase separates between either an empty system or a system at a finite density.

The effect of the nearest-neighbor repulsion is shown in Fig.~\ref{pd}b for $V=2.0$ and $N_{\rm s}=25$ and Fig.~\ref{pd}c for $V=2.8$ and $N_{\rm s}=32$. It competes with the attractive interaction originating from the correlated hopping. The idea is to destabilize the "many-boson" bound states while keeping stable bound states of two bosons. In the limit of large $V$, both PS and PSF are destroyed and a standard SF at low density is realized. Note also the
appearance of yet another region of phase separation around densities of $n=0.5$ for $V=2.8$. This region of
PS is related to the first order transition out of the checkerboard solid reported in Ref.~[\onlinecite{batro00}].
At intermediate repulsions, the situation is more complex. In Fig.~\ref{pd}b for $V=2.0$, the PSF is stabilized for correlated hopping dominating the normal hopping at small densities. Still a region of PS remains between the PSF and the SF, leading to a region of excluded densities between the two stable phases. At the even larger value of $V=2.8$, (Fig.~\ref{pd}c) the PS is completely suppressed, while a reduced region is maintained where the PSF is stable.
\begin{figure}
    \begin{center}
     \includegraphics[width=\columnwidth]{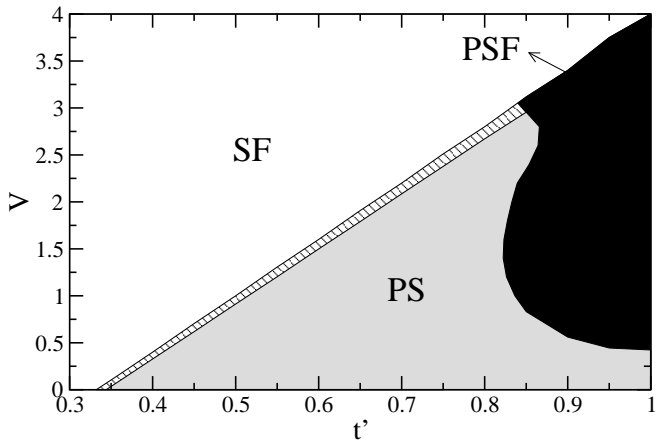}
    \end{center}
    \caption{Extension of the different phase in the low-density limit. White represents the SF phase, grey denotes PS, and black represents the PSF. The tiny shaded region marks a region of pairing which is only present for a finite size system.}
    \label{binding}
\end{figure}

The major outcome of the phase diagram for the finite clusters is that the PSF is indeed realized in a small window of finite nearest-neighbor repulsion and at low densities. In the following we will study the dilute low-density limit in order to see whether we get a consistent physical picture. We therefore calculated the ground state energies $E(1)$, $E(2)$, and $E(4)$ for a $N_{\rm s}=58\ $\footnote{The $N_{\rm s}=58$ cluster is a square sample where the edges are spanned by the vector (7,3)} cluster with one, two and four particles, and form the two and four particle binding energies $\Delta_2=E(2)-2E(1)$ and $\Delta_4=E(4)-2E(2)$. We attribute a bound state of four particles $(\Delta_4<0)$ to a precursor of PS, a bound state of two particles $(\Delta_2<0, \Delta_4>0)$ corresponds to the PSF, and the case of no bound states $(\Delta_2>0, \Delta_4>0)$ corresponds to the SF phase, in analogy to the binding of hole-pairs in the 2D $t{-}J$ model \cite{poilb95}. 
In Fig.~\ref{binding} the results of such an analysis are displayed as a function of $t^\prime$ and $V$. The results are along the lines we have deduced from the calculations at finite density on the smaller clusters above. At zero $V$ there is only a SF phase and PS. But the PSF can be stabilized by finite $V$ at large correlated hopping. For $t=0$, a pair of nearest-neighbor bosons
hops on an effective square lattice with amplitude $t'$, and the pair can gain at most a kinetic
energy of $4t'$ equal to half the band-width. So the boundary at $V=4$ for $t'=1$ is an exact result.

Up to now we have only discussed energies and densities, which provided evidence in favor of 
the presence of a region in the phase diagram where the bosons form pairs. Yet we have not shown that these pairs actually condense. 
Conceptually, one now has to study the asymptotic behavior of the relevant Green's functions, i.e. two-point correlation functions for the single-boson SF and four-point correlation functions for the PSF. An exponential decay of the single-particle Green's function at zero temperature signals short-range two-boson correlations and therefore the absence of a single-boson SF condensate. But how does one deal with the PSF?

\begin{figure}[b]
    \begin{center}
     \includegraphics[width=\columnwidth]{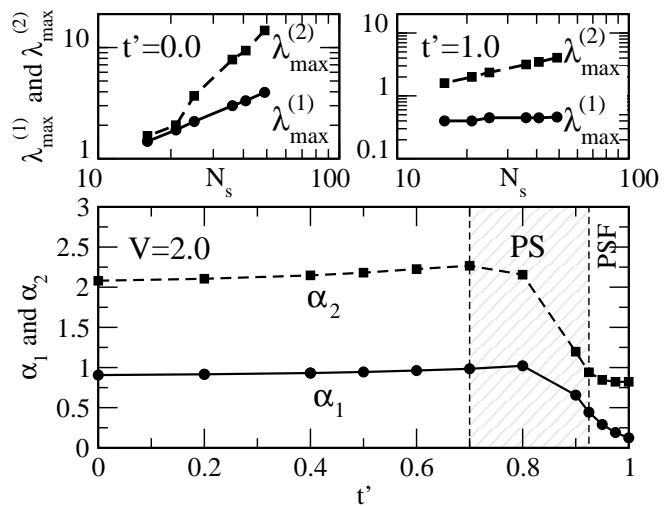}
    \end{center}
    \caption{The maximum eigenvalues $\lambda_{\rm max}^{(1)}$ and $\lambda_{\rm max}^{(2)}$ for $V=2$ and $n=0.1$ as a function of $N_{\rm s}$ for $t^\prime=0$ ({\it upper left panel}) and $t^\prime=1$ ({\it upper right panel}).\\{\it Lower panel}: Exponents $\alpha_{1}$ and $\alpha_{2}$ for $V=2$ as a function of $t^\prime$ for constant density $n=0.1$. The exponent $\alpha_{1}$ is shown as solid curves and the exponent $\alpha_{2}$ is shown as dashed curves.}
    \label{eigen}
\end{figure}
It was realized by Yang\cite{yang62} that the reduced density matrices
\begin{eqnarray}
 \rho^{(1)}_{i,j} &=& \langle b^\dagger_i b^{\phantom{\dagger}}_j\rangle \\
 \rho^{(2)}_{\bar{i},\bar{j}} &=& \langle b^\dagger_{i_1} b^\dagger_{i_2} b^{\phantom{\dagger}}_{j_1}b^{\phantom{\dagger}}_{j_2}\rangle
\end{eqnarray}
are the relevant quantities to investigate in order to answer this question. Here $\rho^{(1)}_{i,j}$ is a $N_{\rm s}\times N_{\rm s}$-matrix with $i,j\in\{1,\ldots,N_{\rm s}\}$ and $\rho^{(2)}_{\bar{i},\bar{j}}$ a $N_{\rm s}^2\times N_{\rm s}^2$-matrix with $\bar{i}=(i_1,i_2)\in\{1,\ldots,N_{\rm s}^2\}$ and $\bar{j}=(j_1,j_2)\in\{1,\ldots,N_{\rm s}^2\}$. The behavior of the largest eigenvalue $\lambda_{\rm max}^{(1)}$ of $\rho^{(1)}$ and $\lambda_{\rm max}^{(2)}$ of $\rho^{(2)}$ as function of the system size $N_{\rm s}$ for constant density is decisive. For the SF phase one finds $\lambda_{\rm max}^{(1)}\propto N_{\rm s}$ and $\lambda_{\rm max}^{(2)}\propto N_{\rm s}^2$. In contrast for the PSF one has $\lambda_{\rm max}^{(1)}\propto 1$ and $\lambda_{\rm max}^{(2)}\propto N_{\rm s}$. The general behavior is summarized in Tab.~\ref{tab_1}. 
\begin{table}
\begin{center}
\begin{tabular}{c|c|c}
  Phase & Eigenvalue $\lambda_{\rm max}^{(1)}$ & Eigenvalue $\lambda_{\rm max}^{(2)}$\\
  \hline\hline
  SF & O$(N_{\rm s})$ & O$(N_{\rm s}^2)$\\
  \hline
  PSF & O$(1)$ & O$(N_{\rm s})$\\
\end{tabular}
\end{center}
\caption{\label{tab_1} General behavior of the maximum eigenvalues $\lambda_{\rm max}^{(1)}$ and $\lambda_{\rm max}^{(2)}$ for the different superfluid phases SF and PSF.}
\end{table}
The different behaviors of $\lambda_{\rm max}^{(1)}$ and $\lambda_{\rm max}^{(2)}$ can be understood as follows. Let us consider first $\rho^{(1)}_{i,j}$. For any phase without a usual single-boson condensate all large distance two-point correlations are generically exponentially suppressed, i.e. all entries in $\rho^{(1)}_{i,j}$ with $|i-j|\gg 1$ are very small and the matrix has only a finite number of sizable entries for fixed $j$. In the limit of no correlations when $i\neq j$, one has a diagonal matrix with maximum eigenvalue of order $1$. In contrast, for a standard SF one has long-range two-point correlation functions and {\it all} entries of $\rho^{(1)}_{i,j}$ are of the same order. For the case where all correlations are exactly the same, one can easily convince oneself that the maximum eigenvalue is of the order $N_{\rm s}$. The behavior of $\rho^{(2)}$ is more complex. For the case of no single-boson {\it and} no boson-pair condensate all four-point correlations are suppressed exponentially and one again obtains in complete analogy a maximum eigenvalue of order $1$. For the standard SF all four-point correlation functions are long-ranged since the four-point correlators contain always two-point correlation functions, i.e. {\it all} entries of $\rho^{(2)}$ are of the same order and the maximum eigenvalue $\lambda_{\rm max}^{(2)}$ is of order $N_{\rm s}^2$. The PSF has suppressed two-point correlation functions and simultaneously long-range correlations between two boson pairs. Therefore all entries involving large distances are basically zero except the ones where two distances $|i_1-j_1|\lesssim \xi$ and $|i_2-j_2|\lesssim \xi$ (or the other combination of indices) are within a pairing length. This gives a maximum eigenvalue $\lambda_{\rm max}^{(2)}$ of order $N_{\rm s}$. In this way one can either have long-range order of single bosons or long-range order of boson pairs. It does not make sense to speak about ordering in both channels simultaneously since a condensate of single bosons is automatically accompanied by a pair condensate.

In order to compare the behavior of $\lambda_{\rm max}^{(1)}$ and $\lambda_{\rm max}^{(2)}$ for the different phases, we concentrate on systems with an even number of bosons at constant and low densities. Since we are limited by the size of the considered systems we use the following procedure to characterize $\lambda_{\rm max}^{(1)}$ and $\lambda_{\rm max}^{(2)}$. We consider the maximum eigenvalues $\lambda_{\rm max}^{(1)}$ and $\lambda_{\rm max}^{(2)}$ as a function of $N_{\rm s}$ and $N_{\rm p}$ where we restrict $N_{\rm p}$ to be even. Using a linear interpolation between the calculated points we obtain the maximum eigenvalues $\lambda_{\rm max}^{(1)}(N_{\rm s},n={\rm const})$ and $\lambda_{\rm max}^{(2)}(N_{\rm s},n={\rm const})$ as a function of $N_{\rm s}$ for a constant density. The resulting data for $V=2$,$n=0.1$, and the extreme cases $t^\prime\in\{0;1\}$ are shown in the upper panels of Fig.~\ref{eigen}. Then we fit the functions $\lambda_{\rm max}^{(1,2)}(N_{\rm s},n={\rm const})$ with $a_{1,2}N_{\rm s}^{\alpha_{1,2}}$. The exponents $\alpha_1$ for $\lambda_{\rm max}^{(1)}(N_{\rm s},n={\rm const})$ and $\alpha_2$ for $\lambda_{\rm max}^{(2)}(N_{\rm s},n={\rm const})$ can then be easily used to characterize the asymptotic behavior of $\lambda_{\rm max}^{(1)}$ and $\lambda_{\rm max}^{(2)}$. The exponents $\alpha_{1}$ and $\alpha_{2}$ are shown in the lower panel of Fig.~\ref{eigen} for $V=2.0$ and as a function of $t^\prime$ for a density $n=0.1$.

The system realizes a SF phase for small values of $t^\prime$. This is nicely reflected in the displayed exponents. $\alpha_{1}$ is close to the expected value of $1$ and $\alpha_{2}$ is close to the value of $2$. Deviations are presumably due to finite size effects. It can be clearly seen that these ratios change drastically in the regime of large correlated hopping. 
The deviations are strongest for $t^\prime\approx 1$. Here one obtains $\alpha_{1}\approx 0$ and $\alpha_{2}\approx 1$, i.e. values which are close to the expected values for the PSF. In the intermediate $t^\prime$-regime where one has PS no particular exponents are expected or observed. The investigation of the maximum eigenvalues of $\rho^{(1)}$ and $\rho^{(2)}$ yields therefore additional evidence for the presence of the PSF for large $t^\prime$ at low densities, i.e.~the system realizes a stable condensate of boson pairs. 

\section{Quantum Monte Carlo}
\label{QMC}
In this section we exploit the results obtained by QMC and we complement the physical picture deduced from the ED results, i.e.~we will establish the existence of the PSF in a finite window for finite nearest-neighbor repulsion. The second part of this section will address the question of possible phase transitions out of the PSF.

The model under study does not suffer from a sign problem and is therefore accessible to quantum Monte Carlo (QMC) techniques. However, the treatment of three-site terms is not standard. We will use a SSE algorithm\cite{sandv91,sandv97,sylju03} to tackle the problem as it has been shown to be able to treat multi-site terms\cite{melko05}. Some details of the algorithm and its implementation are given in the Appendix.

We start the discussion about the SSE results by important comments on the numerical efficiency of the one-worm SSE algorithm for the case of the pairing phase. A basic property of the PSF are exponentially decaying two-point correlation functions. Since typical worm algorithms are based on the insertion of two discontinuities into the system which propagate in an extended configuration space\cite{proko98} (relevant for measuring two-point correlation functions), the worm updates in space and time are short-ranged. Normally, this does not cause a problem because the single-boson SF becomes unstable to insulating phases which have short ranged Green's functions like charge-ordered states. For the PSF this is different. It is a long-range ordered phase with short range two-point correlation functions. The very nature of the PSF itself therefore leads to intrinsic algorithmic problems for single worm updates. Nevertheless, we believe that a certain number of important observations can be made. 

The stiffness of a SF phase is determined in SSE by measuring the winding number\cite{sandv97,pollo87}. If a macroscopic fraction of the particles winds around the system, the system acquires a finite stiffness. The SSE algorithm uses a starting configuration with fixed zero winding number. In the single-boson SF the worm algorithm is efficient and the algorithm is able to let a large number of bosons to wind around the system. In the PSF however,
we generically expect pairs of bosons to wind around the system. This expectation is indeed confirmed by SSE simulations on small systems, where 
winding number histograms show a very clear even-odd effect, i.e. only even winding numbers occur.
But the accumulation of a large number of winding bosons pairs is exponentially suppressed by the short range nature of the single worm updates in the PSF. 
The higher winding numbers and therefore the stiffness are not well explored in the PSF, based on the present worm update scheme. 
We are currently working on two-worm update schemes which should improve the efficiency of the algorithm in the PSF\cite{schmi06}.

In contrast to the winding number, the density of the system can be randomly set in the starting configuration. It is therefore possible to use for a given parameter set a large number of different seeds and average over the different runs. In this way one has access to the total density and also to density histograms. The latter counts how many times a system with a certain total number of bosons is realized. One expects a pronounced even/odd effect in the PSF which should be absent for a standard SF.

The third quantity we focus on is the equal-time one-particle Green's function $G(r,\tau=0)=\langle b^\dagger_r b^{\phantom{\dagger}}_0\rangle$. We expect that these correlations are exponentially decaying in the PSF, due to the pairing gap, while they are long-ranged in the conventional SF. 

Based on the ED results we expect that the PSF is realized at low densities for large correlated hopping and at corresponding values of the nearest-neighbor repulsion. In the following, we will concentrate on two parameter sets: $t=0.1$, $t^\prime =0.9$, and $V=2.0$ on one hand, and $t=0.05$, $t^\prime =0.95$, and $V=2.8$ on the other hand. In the first case we expect the PSF at low densities to be separated from the standard SF by an excluded density region of PS. 
For the second parameter set ED results predict the PSF at low densities with possibly a direct transition to the standard SF. 
In the following we always investigate the case $T=0.05$

\begin{figure}
    \begin{center}
     \includegraphics[width=\columnwidth]{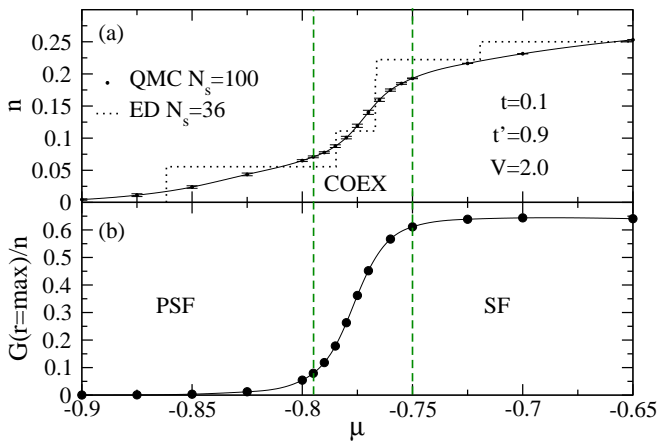}
    \end{center}
    \caption{(a) Density as a function of chemical potential for $T=0.05$, $t=0.1$, $t^\prime=0.9$, $V=2.0$ obtained by QMC for $N_{\rm s}=100$. Error bars are one standard deviation. Number of seeds is 16. The dotted black line denote ED results with $N_{\rm s}=36$. (b) The ratio $G(r={\rm max})/n$ with $r=\sqrt{50}$ as a function of the chemical potential for the same parameter set giving a measure for the one-particle condensate. The region between the two grey lines (guide to the eyes) signals the finite-temperature coexistence region (COEX).}
    \label{nvsmu_1}
\end{figure}
\begin{figure}
    \begin{center}
     \includegraphics[width=\columnwidth]{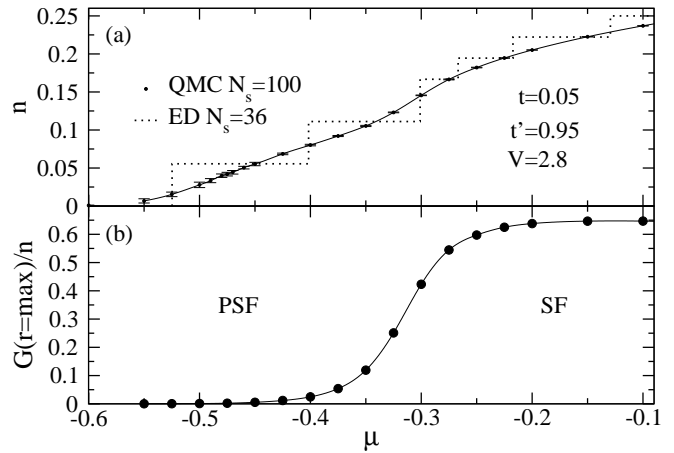}
    \end{center}
    \caption{(a) Density as a function of chemical potential for $T=0.05$, $t=0.05$, $t^\prime=0.95$, $V=2.8$ obtained by QMC for $N_{\rm s}=100$. Error bars are one standard deviation. The number of seeds is 32. The dotted black line denote ED results with $N_{\rm s}=36$. (b) Ratio $G(r={\rm max})/n$ with $r=\sqrt{50}$ as a function of the chemical potential for the same parameter set giving a measure for the one-particle condensate.}
    \label{nvsmu_2}
\end{figure}

Since the algorithm is grand canonical, we  first take a look at the total density as a function of the chemical potential for both parameter sets in order to determine the low-density region of the system. This is shown in Fig.~\ref{nvsmu_1}a and in Fig.~\ref{nvsmu_2}a for a $N_{\rm s}=100$ system. The dotted lines represent the ED results at zero temperature for the $N_{\rm s}=36$ cluster. We expect from the ED results the PSF to be stable for $n<0.11$ ($\mu<-0.77$) for the first set and for $n<0.17$ ($\mu<-0.27$) for the second set of parameters. Clearly, both parameter sets show no plateaux in $n(\mu)$, ruling out the realization of an insulating phase.  

\begin{figure}
    \begin{center}
     \includegraphics[width=\columnwidth]{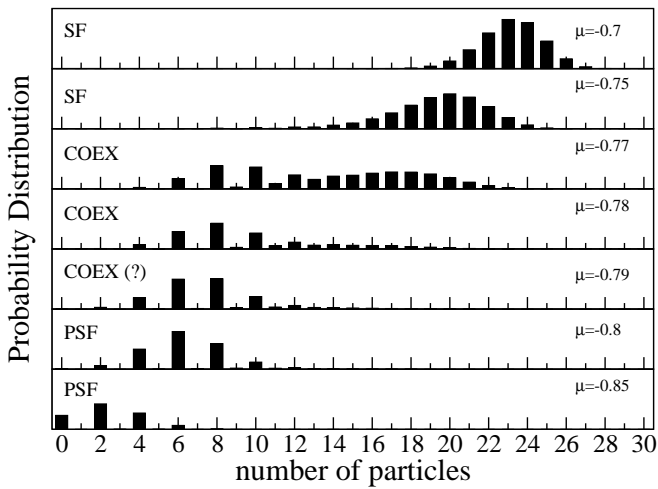}
    \end{center}
    \caption{Density histogram for $N_{\rm s}=100$, $T=0.05$, $t=0.1$, $t^\prime=0.9$, $V=2.0$ and different values for the chemical potential. Error bars are one standard deviation.}
    \label{histo_n_1}
\end{figure}

\begin{figure}
    \begin{center}
     \includegraphics[width=\columnwidth]{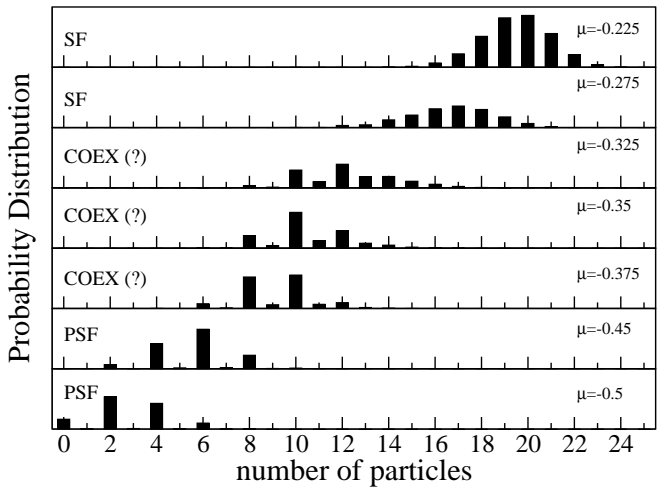}
    \end{center}
    \caption{Density histogram for $N_{\rm s}=100$, $T=0.05$, $t=0.05$, $t^\prime=0.95$, $V=2.8$ and different values for the chemical potential. Error bars are one standard deviation.}
    \label{histo_n_2}
\end{figure}

For a given system size, a convenient measure of the presence of a one-particle condensate is given by the ratio
of the one-particle Green's function at the maximal distance allowed by the cluster
to the density $G(r={\rm max})/n$. This ratio is plotted in Fig.~\ref{nvsmu_1}b and Fig.~\ref{nvsmu_2}b. For large chemical potential this quantity is finite and the system is in a standard SF phase. On the contrary, if this quantity vanishes it signals the breakdown of the SF. The finite region between these two limits marks the finite-temperature coexistence region (COEX). 
Clearly, the SF is unstable for both parameter sets at low densities. The corresponding one-particle Green's functions $G(r)$ show a strong decay with distance of the two-particle correlation functions in this density regime. But the phase present at low densities has to be still characterized since it could in principle be either the PSF or PS. To this end, the density histogram is displayed in Fig.~\ref{histo_n_1} and Fig.~\ref{histo_n_2}. It is expected that this histogram is fundamentally different for the two phases. In the PSF one gaussian with a large odd/even effect should be present while one expects a superposition of two gaussians (one reflecting a subsystem realizing a SF phase and a second one reflecting a PSF) for a PS phase. Focusing only on low densities (small values of the chemical potential) the histograms for both parameter sets display one gaussian with almost only contributions with an even number of particles. This establishes that at low densities the PSF is a stable phase.

Let us briefly summarize our characterization of the large $t'$, low density phase. 
Whenever we have only a fluctuating even number of bosons in the system and when we observe simultaneously exponentially decaying one-particle correlations, it is natural to conjecture that we
are in presence of a paired superfluid phase. 
Of course, on the basis of these Quantum Monte Carlo results alone, we cannot exclude the possibility 
of a paired phase with vanishing stiffness since we presently have no direct access to this quantity, 
but this seems unlikely on the basis of the results of the previous section: The size scaling 
$\lambda_{\rm max}^{(2)}\propto N_{\rm s}$ strongly suggests pair off-diagonal long-range order.
Besides, an insulating behaviour can be excluded since it would be easily detected as plateaux 
in the density as a function of the chemical potential, which are clearly not present in the Quantum
Monte Carlo data. So altogether the existence of a PSF seems to be on solid grounds.

A second important issue is the nature of possible quantum phase transitions out of such an unconventional PSF. We remind the reader that the two superfluid phases (SF and PSF) have distinct symmetries enabling the possibility of a true quantum phase transition between them. So the transition between the PSF and the SF can 
be either a first order transition,  or a second order transition in the Ising universality 
class\cite{bendj05,kagan02,kuklo04,roman04,radzi04}. The available 
data are not sufficient to answer this question in a rigorous fashion. This is mainly due to the above-mentioned intrinsic algorithmic problems of the worm updates. Besides, to decide between a weakly first order 
and a second order transition is  numerically hard. Nevertheless, important trends can be deduced.

Let us first discuss the case $t=0.1$, $t^\prime=0.9$, and $V=2.0$. Results for this parameter set are shown in Fig.~\ref{nvsmu_1} and Fig.~\ref{histo_n_1}. ED predicts a first order transition between the two superfluid phases. A first order transition at zero temperature corresponds to a jump of the total density as a function of the chemical potential. The finite temperature nature of the SSE results will smear out this jump slightly. Indeed, the SSE data in Fig.~\ref{nvsmu_1} display a rather sharp dependence of the density as function of the chemical potential close to $\mu=-0.77$ as expected from the ED data. Additionally, the density histogram in Fig.~\ref{histo_n_1} displays a finite-temperature coexistence region (COEX) of the typical behavior of both superfluids near $\mu=-0.77$: There is a gaussian with only even particle number at lower densities characteristic for the PSF and a second gaussian with even and odd particle number at higher density representing a standard single-boson SF. It is therefore likely that the transition for this parameter set is first order.

ED predicts a continuous transition for the second parameter set $t=0.05$, $t^\prime=0.95$, and $V=2.8$. The corresponding results are shown in Fig.~\ref{nvsmu_1} and Fig.~\ref{histo_n_1}. Indeed, the density as a function of the chemical potential displays a much smoother behavior as shown in Fig.~\ref{nvsmu_2}. A clear jump is not visible but one can identify an inflection point close to $\mu=-0.33$ signaling a change of phase. The same trend can be seen in the density histogram in Fig.~\ref{histo_n_2}. There is a rather smooth evolution of the density histogram from the PSF to the SF. But the numerical data are not sufficient to decide whether a weakly first order (existence of a tiny coexistence region) or a second order transition between the two superfluids is present. It is likely that the second order transition is only realized (if at all) in a narrow window for large enough values of the nearest-neighbor repulsion.
\section{Conclusion}
\label{Conclusion}
In this work we studied a model of hard-core bosons on the square lattice with correlated hopping. We used ED and SSE simulations to establish the existence of an unconventional phase of condensed boson pairs at zero temperature. In the PSF two-point correlations are short ranged while there is long-range order in the four-point correlations. In other words the bosons form stable molecular pairs which condense. We found that a finite nearest-neighbor repulsion is crucial for such a phase. The correlated hopping acts as an attractive force between the bosons. As a consequence, the system is unstable against PS for small nearest-neighbor repulsion. A macroscopic part of the bosons glue together. In the opposite limit of large repulsion all bound states between the bosons are destroyed and a standard SF is realized. But the major outcome of this work is that there is a finite window of nearest-neigbour repulsion where the formation of a condensate of paired bosons is stabilized. The nearest-neighbor repulsion is large enough to destroy the PS but it is small enough not to destroy the boson pairs.

A second important aspect is the nature of the phase transition out of the PSF. This can be either of first or second order. Clearly, the realization of a continuous transition would be of great interest since it is expected to be in the Ising universality class. The detailed nature of the transition at zero but also at finite temperature is an open issue and can be studied microscopically in our model with correlated hopping. The numerical efficiency of the algorithm we have used is currently not sufficient to answer the question of the nature of the transition rigorously. But it is likely that the continuous transition is only realized at large values of the nearest-neighbor repulsion. When the repulsion is too small, the transition is probably first order.

In conclusion, we have established in this work the existence of an unconventional pairing phase of hard-core bosons in a microscopic model with correlated hopping. We have seen that current SSE simulations based on worm updates are not efficient in a phase of boson-pair condensation. The development of updates schemes bases on two worms are currently under study\cite{schmi06} and should enable the clarification of the nature of the phase transitions out of this exotic phase at zero and at finite temperature.  
\begin{acknowledgments}
We thank F. Alet, R. Bendjama, H.P. B\"uchler, B. Kumar and L. Pollet for stimulating discussions. The SSE simulations were done using the ALPS libraries\cite{alps, alet05}. We also acknowledge the Swiss National Funds and the MaNEP for financial support and the CSCS (Manno)
for allocation of computing time on the Cray XT3.
\end{acknowledgments}

\appendix*
\section{SSE implementation}
In this appendix we mention some technical points allowing the reader to follow the numerical implementation of the problem. For some introductory literature on SSE we refer to Ref.~\cite{sandv99,sylju02}. The SSE simulations were done using a modified
version of the ALPS SSE code~\cite{alet05} based on the ALPS libraries\cite{alps}. Conventionally, SSE algorithms use bond terms as basic building blocks. The presence of three-site terms demands a larger building block. We have chosen as basic objects of our problem operators acting on a plaquette of the square lattice\cite{melko05}. In the following we denote the four sites of one plaquette $i$, $j$, $k$, and $l$ in anti-clockwise direction putting the first site in the left lower corner. We have three classes of vertices. First, there are diagonal vertices
\begin{eqnarray}
 H_{\rm 1,a}&=&CI_{ijkl}\\\nonumber
 &&+\frac{\mu}{4}\left[ n_i I_{jkl}+n_j I_{ikl}+n_k I_{ijl}+n_l I_{ijk}\right]\\\nonumber
 &&-\frac{V}{2}\left[ n_i n_j I_{kl}+ n_j n_k I_{il}+ n_k n_l I_{ij}+ n_l n_i I_{kj}\right]\quad . 
\end{eqnarray}
Here $C$ is a constant which has to be adjusted such that all diagonal contributions are not negative, i.e. that there is no sign problem. Second, there are bond vertices
\begin{eqnarray}
 H_{\rm 2,a}&=&\frac{t}{2}B_{ij}I_{lk}\\
 H_{\rm 3,a}&=&\frac{t}{2}B_{jk}I_{il}\\
 H_{\rm 4,a}&=&\frac{t}{2}B_{kl}I_{ij}\\
 H_{\rm 5,a}&=&\frac{t}{2}B_{li}I_{jk}
\end{eqnarray}
where $B_{ij}=b^\dagger_j b^{\phantom{\dagger}}_i+b^\dagger_i b^{\phantom{\dagger}}_j$. Third, we have vertices originating from the correlated hopping
\begin{eqnarray}
 H_{\rm 6,a}&=&t^\prime T_{ilj}I_{k}\\\nonumber
            &=&t^\prime I_k n_i B_{lj}\\
 H_{\rm 7,a}&=&t^\prime T_{jik}I_{l}\\\nonumber
            &=&t^\prime I_l n_j B_{ik}\\
 H_{\rm 8,a}&=&t^\prime T_{klj}I_{i}\\\nonumber
            &=&t^\prime I_i n_k B_{lj}\\
 H_{\rm 9,a}&=&t^\prime T_{lik}I_{j}\\\nonumber
            &=&t^\prime I_j n_l B_{ik} \quad .
\end{eqnarray}
A second important issue is the choice of the updates. Beside standard diagonal updates we use directed loop updates. The determination of the travel probabilities of the two discontinuities in order to fulfill detailed balance for the created directed loops is a non-trivial problem. A successfull strategy is to minimize the bounce probabilities having in mind that one wants to create large directed loops. Following Ref.~\cite{sylju03}, one can write down a solution which minimizes the bounce amplitudes. The bounce processes are only non-zero in a parameter regime where one term in the Hamiltonian dominates the other terms. In all other cases, the maximum weight does not dominates the other weights, and bounce free solutions can be obtained\cite{sylju03}.

\end{document}